\documentclass{aa}

\usepackage[T1]{fontenc}

\DeclareRobustCommand{\VAN}[3]{#2}
\let\VANthebibliography\thebibliography
\def\thebibliography{\DeclareRobustCommand{\VAN}[3]{##3}\VANthebibliography}

\usepackage{graphicx}	
\usepackage{subfigure}
\usepackage{placeins}    
\usepackage{siunitx}		
\usepackage{amsmath}	
\usepackage{amssymb}	
\usepackage{aas_macros}
\usepackage{enumitem} 
\usepackage[shortcuts,acronyms]{glossaries-extra}		
\usepackage{orcidlink}

\newabbr{GC}{GC}{globular cluster}
\newabbr[longplural = core-collapse supernovae, shortplural = CCSNe]{CCSN}{CCSN}{core-collapse supernova}
\newabbr{IMF}{IMF}{initial mass function}
\newabbr{SF}{SF}{star formation}
\newabbr{SFE}{SFE}{star formation efficiency}
\newabbr{SFR}{SFR}{star formation rate}
\newabbr{MW}{MW}{Milky Way}
\newabbr{AGB}{AGB}{asymptotic giant branch}
\newabbr{SMBH}{SMBH}{super-massive black hole}
\newabbr{BH}{BH}{black hole}
\newabbr{NS}{NS}{neutron star}
\newabbr{SFDE}{\textsc{SFDE}}{\textsc{Star Formation Duration Estimator}}

\begin{document}

\title{Failed supernova explosions increase the duration of star formation in globular clusters}

\author{
Henriette Wirth\orcidlink{0000-0003-1258-3162}$^{1}$\thanks{E-mail: wirth@sirrah.troja.mff.cuni.cz (HW)},
Jaroslav Haas$^{1}$,
Ladislav Šubr\orcidlink{0000-0003-1924-8834}$^1$,
Tereza Jerabkova\orcidlink{0000-0002-1251-9905}$^{2}$,
Zhiqiang Yan\orcidlink{0000-0001-7395-1198}$^{3}$
and Pavel Kroupa\orcidlink{0000-0002-7301-3377}$^{1,4}$}
\institute{
$^{1}$Charles University, Faculty of Mathematics and Physics, Astronomical Institute, V Hole\v{s}ovi\v{c}kách 2, Praha, CZ-18000, Czech Republic\\
$^{2}$European Southern Observatory, Karl-Schwarzschild-Strasse 2, 85748 Garching bei M\"unchen, Germany\\
$^{3}$School of Astronomy and Space Science, Nanjing University, Nanjing, 210023, China\\
$^{4}$Helmholtz Institut für Strahlen und Kernphysik, Universität Bonn, Nussallee 1416, 53115 Bonn, Germany\\
}

\date{Accepted XXX. Received YYY; in original form ZZZ}

\abstract
   {The duration of \ac{SF} in \acp{GC} is an essential aspect for understanding their formation.
Contrary to previous presumptions that all stars above 8 $M_\odot$ explode as \acp{CCSN}, recent evidence suggests a more complex scenario.}
   {We analyse iron spread observations from 55 \acp{GC} to estimate the number of \acp{CCSN} explosions before \ac{SF} termination, thereby determining the \ac{SF} duration.
This work for the first time takes the possibility of failed \acp{CCSN} into account, when estimating the \ac{SF} duration.}
   {Two scenarios are considered: one where all stars explode as \acp{CCSN} and another where only stars below 20 $M_\odot$ lead to \acp{CCSN}, as most \ac{CCSN} models predict that no failed \acp{CCSN} happen below 20 $M_\odot$.}
   {This establishes a lower ($\approx 3.5 ~\rm Myr$) and an upper ($\approx 10.5 ~\rm Myr$) limit for the duration of \ac{SF}.
Extending the findings of our previous paper, this study indicates a significant difference in \ac{SF} duration based on \ac{CCSN} outcomes, with failed \acp{CCSN} extending \ac{SF} by up to a factor of three.
Additionally, a new code is introduced to compute the \ac{SF} duration for a given \ac{CCSN} model.}
   {The extended \ac{SF} has important implications on \ac{GC} formation, including enhanced pollution from stellar winds and increased binary star encounters.
These results underscore the need for a refined understanding of \acp{CCSN} in estimating \ac{SF} durations and the formation of multiple stellar populations in \acp{GC}.}


\titlerunning{failed CCSNe in GCs}
\authorrunning{Wirth, Haas, Šubr, Jerabkova, Yan, Kroupa}

\maketitle

\glsresetall



\section{Introduction}
\label{sec_intr}

Knowing how long \ac{SF} lasted is a further piece of the puzzle towards understanding the formation of \acp{GC} and the presence of multiple populations therein \citep{2015MNRAS.450..815M,2015MNRAS.446.1672M,2015MNRAS.450.3750M,2015ApJ...808...51M,2016MmSAI..87..303M,2017MNRAS.464.3636M,2018ApJ...859...81M,2019MNRAS.487.3815M}.
Stellar clusters form rapidly ($\lesssim 10 ~\rm Myr$) at the centres of massive gas cloud cores.
After the first stars are formed the gas is quickly expelled inhibiting the formation of any further stars \citep{1980ApJ...235..986H,1984ApJ...285..141L,2001MNRAS.321..699K,2011MNRAS.415.3439D,2013MNRAS.436.3727D,2015ApJ...814L..14C,2018ASSL..424..143B,2023ApJ...957...77P}.

In \citet{2021MNRAS.506.4131W} and \citet{2022MNRAS.516.3342W} the duration of \ac{SF} was estimated using the observed iron spreads in 55 \acp{GC} taken from \citet{2019ApJS..245....5B}.
For all \acp{GC} the initial masses were computed using a method based on \citet{2003MNRAS.340..227B}, but applying an invariant - in \citet{2021MNRAS.506.4131W} - and systematically varying stellar \ac{IMF} - in \citet{2022MNRAS.516.3342W} - to each \ac{GC}.
From the initial masses of the \acp{GC}, the masses of the initial gas cloud the \acp{GC} formed out of was computed and combining this with the observed iron spreads the overall amount of iron that had to be produced to explain the observed iron spread was computed.
Since \acp{CCSN} are the earliest source of iron in \acp{GC} this allowed \citet{2021MNRAS.506.4131W} and \citet{2022MNRAS.516.3342W} to compute the number of \acp{CCSN} that must have exploded before the end of \ac{SF}.
The progenitor's lifetimes then provide an estimate for the duration of \ac{SF}.

While previous studies exploring the gas expulsion and the end of \ac{SF} assumed that all stars above $8 ~M_\odot$ explode as \acp{CCSN} \citep{2015ApJ...814L..14C,2021MNRAS.506.4131W,2022MNRAS.516.3342W}, newer findings show that especially in the high mass range ($> 30 ~M_\odot$) stars often fail to explode and collapse into a black hole as a so called `failed \ac{CCSN}'.
\citet{2003ApJ...591..288H,2011ApJ...730...70O,2015ApJ...801...90P,2016ApJ...821...38S,2019ApJ...870....1E} and \citet{2020rfma.book..189P} studied the nature of \acp{CCSN} up to a mass of $120 ~M_\odot$ theoretically.
While \citet{2003ApJ...591..288H} approximates the outcomes of massive stars dying based on a number of previous studies over a wide range of masses and metallicities, the other studies listed here computed smaller samples with more detailed stellar evolution.
\citet{2011ApJ...730...70O} and \citet{2015ApJ...801...90P} investigated models of zero, $10^{-4}$ times solar and solar metallicity; \citet{2016ApJ...821...38S} and \citet{2019ApJ...870....1E} only looked at models with solar metallicity.
While these studies yield vastly different results, when it comes to the outcomes of the evolution of individual stars, they all show that the majority of stars with initial masses up to $20 ~M_\odot$ explode, while stars of higher masses are more likely to end their lives as failed \acp{CCSN}.
It is important to underline here that these are statistical tendencies, rather than strict rules.

The present work, therefore, investigates what effect these failed \acp{CCSN} have on estimations of a \ac{GC}'s \ac{SF} duration from measured iron spreads.
Building on the results of \citet{2022MNRAS.516.3342W} the duration of \ac{SF} is computed assuming all stars more massive than $20 ~M_\odot$ end as failed \acp{CCSN}, below which no failed \acp{CCSN} are expected.
This results in an upper limit for the duration of \ac{SF}.
Despite a large amount of modelling attempts, the exact iron output and whether or not a star results in a \ac{CCSN} based on its metallicity and mass is currently unknown \citep{2003ApJ...591..288H,2011ApJ...730...70O,2015ApJ...801...90P,2016ApJ...821...38S,2019ApJ...870....1E,2020rfma.book..189P}.
Therefore, an open source code named \ac{SFDE} is introduced in Sec. \ref{sec_SFDE} which can take different \ac{CCSN} models as input parameters.
The usage of this code is demonstrated using one specific example.

\section{Methods}
\label{sec_methods}

\subsection{Model assumptions}
\label{sec_ass}

\begin{figure}
    \includegraphics[scale=1]{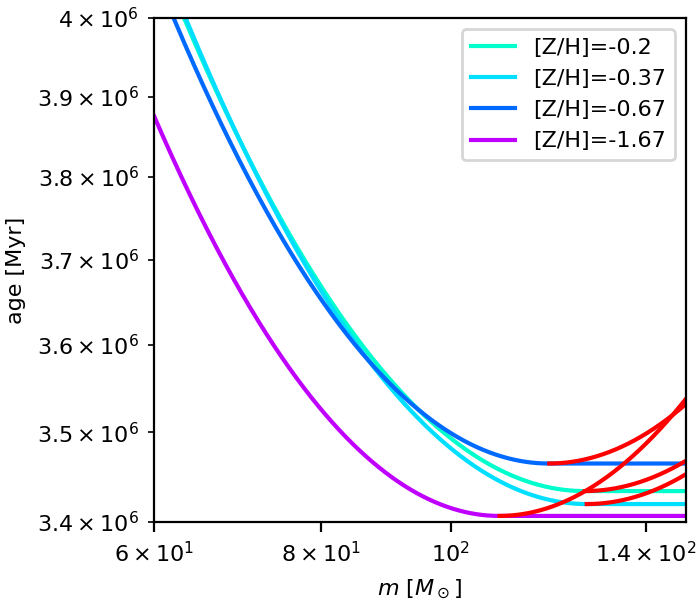}
    \caption{Stellar lifetimes over the stellar masses of high-mass stars according to \cite[][see their fig. 3]{2019A&A...629A..93Y}.
    The areas in which stellar lifetimes increase with initial stellar mass are marked in red. Instead of following these increases, the stellar lifetimes are kept constant beyond the minimum as shown on the colours of the different graphs.}
    \label{fig_stellarlife}
\end{figure}

Following \citet{2022MNRAS.516.3342W} the following assumptions are made:
\begin{enumerate}[leftmargin=1.4em,itemindent=-0.0em]
    \item Continuous \ac{SF} from a gas clump with a \ac{SFE} of 0.3 is assumed.
    This clump mass is linearly correlated to the amount of iron that needs to be produced to explain the iron spread \citep[see][]{2021MNRAS.506.4131W}.
    At the onset of \ac{SF}, the gas has an iron abundance of $[Fe/H] - \sigma_{[Fe/H]}$ \citep{2021MNRAS.506.4131W}, where $[Fe/H]$ is the mean iron abundance of the \ac{GC} and $\sigma_{[Fe/H]}$ the observed iron spread.
    \ac{SF} continues until the gas cloud reaches an iron abundance of $[Fe/H] + \sigma_{[Fe/H]}$.
    The \ac{CCSN} contributing the last increment of the iron abundance terminates further \ac{SF}.
    The exact star formation history is currently undetermined.
    For this work the simple difference between the initial and final iron abundance of the gas cloud is used.
    This would correspond to a symmetric distribution of the \ac{SFE} over the iron abundance in the gas cloud centred around $[Fe/H]$, which would match the Gaussian fit by \citet{2019ApJS..245....5B} over the observed sample of stars.
    \item Each star which explodes ejects $0.074 ~M_\odot$ of iron \citep{2017ApJ...848...25M}.
    \item All ejecta are preserved in the gas cloud.
    It should be noted here that both the assumption that all gas is preserved in the \ac{GC} and the assumption that the ejecta are preserved are strong simplifications.
    In reality the gas would slowly be removed from the cluster due to stellar winds and \acp{CCSN} and the \acp{CCSN} would carve tunnels through which their ejecta might escape preferentially \citep{2015ApJ...814L..14C}.
    The present work, therefore, provides rough estimates rather than detailed calculations. These are in any case not possible as a \ac{GC}-scale star formation process is currently not computable on a star-by-star hydrodynamical with feedback basis.
    \item The gas in the \ac{GC} is well mixed.
    \item The iron abundance of stars is fixed once formed, which means that we can not increase the iron abundance of stars through accretion.
    \item Any kind of binary evolution is neglected and all stars are treated as single stars.
    \item In \citet{2022MNRAS.516.3342W} only the stellar lifetimes and remnant masses for $[Z/H] = -1.67$ from fig. 3 in \citet{2019A&A...629A..93Y} were used to compute the initial \ac{GC} masses, $M_{\rm ini}$, since the metallicity does not have a significant effect on stellar lifetime for high-mass stars.
    However, the remnant masses vary a lot \citep{2019A&A...629A..93Y}.
    For the present work the algorithm was upgraded to include metallicity dependent remnant masses by interpolating linearly between the different graphs from fig. 3 in \citet{2019A&A...629A..93Y}.
    To compute the initial masses of the \acp{GC} taking stellar and dynamical evolution into account, the algorithm presented in \citet{2022MNRAS.516.3342W} is used.
    This algorithm is based on N-body calculations from \citet{2003MNRAS.340..227B}.
    As the remnant masses vary over about an order of magnitude with the metallicity this significantly changes the amount of mass lost due to stellar evolution at the beginning of \ac{GC} evolution and therefore changes the initial mass estimates.
    \label{item_stellarLife}
    \item One of the main equations the computation of the initial mass, $M_\mathrm{ini}$, is based on is eq. 10 of \citet{2003MNRAS.340..227B}:
    \begin{align}
        \label{eq_tsf}
        \frac{T_\mathrm{diss}}{\mathrm{Myr}} &= \beta \left[ \frac{N}{\ln( 0.02 N )} \right]^x \frac{R_\mathrm{G}}{\mathrm{kpc}} \left( \frac{V_\mathrm{G}}{220 \mathrm{km~ s^{-1}}} \right)^{-1} (1 - \epsilon),
    \end{align}
    with the life- or dissolution-time of the \ac{GC}, $T_\mathrm{diss}$, the initial number of stars, $N$, the distance of the apocentre from the Galactic centre, $R_\mathrm{G}$, the velocity with which the \ac{GC} revolves around the Galactic centre, $V_\mathrm{G}$ and the eccentricity of the \ac{GC}'s orbit, $\epsilon$.
    $\beta$ and $x$ are constants that depend on the King concentration parameter, $W_0$.
    \citet{2003MNRAS.340..227B} give values for $\beta$ and $x$ for $W_0 = 5.0$ and $W_0 = 7.0$.
    In \citet{2022MNRAS.516.3342W} we therefore performed a linear fit through those two values.
    This works well for $x$, which does not vary much.
    $\beta$ was fitted with $\beta = 4.11 - 0.44 W_0$.
    This becomes negative at $W_0 = 9.34$, which would lead to a negative dissolution time and therefore become unphysical.
    What would be expected is that the dissolution time goes towards 0 as the concentration parameter goes towards infinity.
    Therefore, we replace the linear fit of $\beta (W_0)$ with:
    \begin{align}
        \beta = a W_0^{-\mu},
        \label{eq_beta}
    \end{align}
    with the fitting constants $a = 36.63$ and $\mu = 1.835$.
    This function naturally fulfils the boundary condition $\lim\limits_{W_0 \to \infty} \beta = 0$.
    \label{item_compMini}
    \item In \citet{2022MNRAS.516.3342W} the spline fitted stellar lifetimes from fig 3 in \citet{2019A&A...629A..93Y} were used.
    However, the fitting method leads to an increase of the stellar lifetimes with mass at the high-mass end (see Fig. \ref{fig_stellarlife}).
    In the present work this is corrected for by keeping the stellar lifetimes constant with mass after their minimum.
\end{enumerate}

While especially the items \ref{item_stellarLife} and \ref{item_compMini} improve the accuracy of our calculations significantly, this work focuses on another major error source of \citet{2022MNRAS.516.3342W}: the unknown number of failed \acp{CCSN}.
As visible in fig. 3 in \citet{2019A&A...629A..93Y} the time until a star dies does not vary much at high masses.
In \citet{2022MNRAS.516.3342W} the number of \acp{CCSN} that explode before the end of \ac{SF} is much lower than the number of stars more massive than $8 ~M_\odot$ in the \ac{GC} (see their table 1).
In fact for most of them it is much lower than the number of stars more massive than $20~M_\odot$.
This means that even if the other effects, like errors in the iron measurements or inaccuracies in the computation of $M_{\rm ini}$, were to change the number of \acp{CCSN} required by a factor of 10, the \ac{SF} duration would only change marginally.
This is especially true if the required number of \acp{CCSN} decreases.
The existence of a large fraction of failed \acp{CCSN} on the other hand means that a large portion of the function shown in fig. 3 of \citet{2019A&A...629A..93Y} is simply skipped, leaving only the longer-lived lower-mass O stars as possible polluters.
This, as is further discussed in this work, can lead to an appreciable jump in the duration of \ac{SF}.

\subsection{Computing the duration of \ac{SF}}

As shown in \citet{2021MNRAS.506.4131W}, the iron spread observed in \acp{GC} can be used to estimate the number of \acp{CCSN} that must have exploded before star formation ended.
From this the time after which \ac{SF} must have ended was computed assuming all stars with masses above $8 ~M_\odot$ explode as \acp{CCSN}.
Since newer studies have shown that a large portion of massive stars end their life as a failed \ac{CCSN} such an approach provides a lower limit for the duration of \ac{SF}.
The current paper aims to find an upper limit by investigating how long \ac{SF} would have to last if only stars with masses below $20 ~M_\odot$ contribute to the iron enrichment.
The value of $20 ~M_\odot$ is chosen, since according to \citet{2011ApJ...730...70O} no failed \acp{CCSN} occur below this mass and according to \citet{2015ApJ...801...90P} and \citet{2016ApJ...821...38S} the overwhelming majority of stars between 8 and $20 ~M_\odot$ explode as \acp{CCSN}.

The present work is based on \citet{2022MNRAS.516.3342W}, which uses an empirically gauged \ac{IMF} as described originally in \citet{2012MNRAS.422.2246M} and updated by \citet{2021A&A...655A..19Y}.
This \ac{IMF} is a function of the initial gas cloud density and metallicity.
Under the assumption that initially more massive stars have shorter lifetimes, the number of \acp{CCSN} required to explain the iron spread can be used to compute the duration of \ac{SF} in \acp{GC}.
The number of required \acp{CCSN}, $N_\mathrm{SN}$, computed from the iron spread, needs to be the result of an integral over the \ac{IMF}, $\xi(m)$, from an unknown minimum value, $m_\mathrm{last}$, to the most massive star to explode with a mass of $20 ~M_\odot$ for the upper limit of the duration of \ac{SF} ($150 ~M_\odot$ is used for the lower limit of the duration of \ac{SF}):
\begin{align}
    N_\mathrm{SN} &= \int\limits_{m_\mathrm{last}}^{20 ~M_\odot} \mathrm{d}m ~\xi (m),\\
    m_\mathrm{last} &= \left( \left( 20 ~M_\odot \right)^{1 - \alpha_3} - N_\mathrm{SN} \frac{1 - \alpha_3}{k_3} \right)^{\frac{1}{1 - \alpha_3}},
\end{align}
where $\alpha_3$ and $k_3$ are the parameters of the \ac{IMF}, $\xi(m) = k_i m^{-\alpha_i}$, for stellar initial masses, $m$, above $1 ~M_\odot$.
As mentioned above, these are functions of the metallicity and the density of the star-forming gas cloud as given in \citet{2022MNRAS.516.3342W}.

The lifetime of a star with mass $m_\mathrm{last}$ then equals the time for which \ac{SF} lasts.
As in \citet{2022MNRAS.516.3342W} the values from fig. 3 in \citet{2019A&A...629A..93Y}, showing the life expectancy of a star over its initial mass, are used to compute the time at which \ac{SF} ends from $m_\mathrm{last}$.
For details of the calculations the reader is referred to \citet{2022MNRAS.516.3342W}.

\subsection{A program for detailed simulation}
\label{sec_SFDE}

While a rough upper and lower estimate can be computed using the methods above, it is possible to do more accurate computations if concrete \ac{CCSN} models are given.
In this section, we therefore present the code \ac{SFDE} that can compute \ac{SF} durations from given tables for iron ejecta and the \ac{CCSN} status (failed or exploded) depending on the stellar mass.
This code computes the amount of iron to be produced as described above and then goes through the stars of the \ac{GC} from the most massive to the least massive one.
In Sec. \ref{sec_ass} the assumption that each star ejects $0.074 M_\odot$ of iron upon explosion was mentioned.
In \ac{SFDE} the amount of iron ejected per star is given in an input file that allows the user to define initial stellar mass, $m$, dependent ejecta (information on input and output files are available in the README.md of \ac{SFDE}).
For each star that explodes the amount of iron is looked up from the table in the given input file and subtracted from the total amount of iron needed.
If the amount of iron still required reaches 0, the last star contributing to \ac{SF} was found.
The life expectancy of this star equals the time for which \ac{SF} lasts in the \ac{GC}.
An open source version of the code \ac{SFDE} can be found on github\footnote{\url{https://github.com/Henri-astro/SFDE}}.
For the present work version 1.0.0 is used.

\section{Results}
\label{sec_res}

\subsection{Changes to the computed initial masses}

\begin{figure}
    \includegraphics[scale=1]{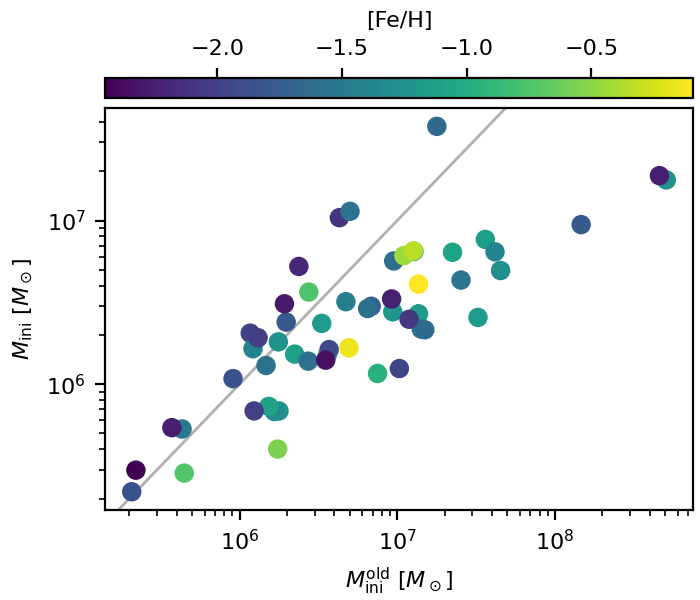}
    \caption{New initial GC masses over the old ones colour-coded by metallicity.
    The identity is shown in grey.}
    \label{fig_Mini}
\end{figure}

\begin{figure}
    \includegraphics[scale=1]{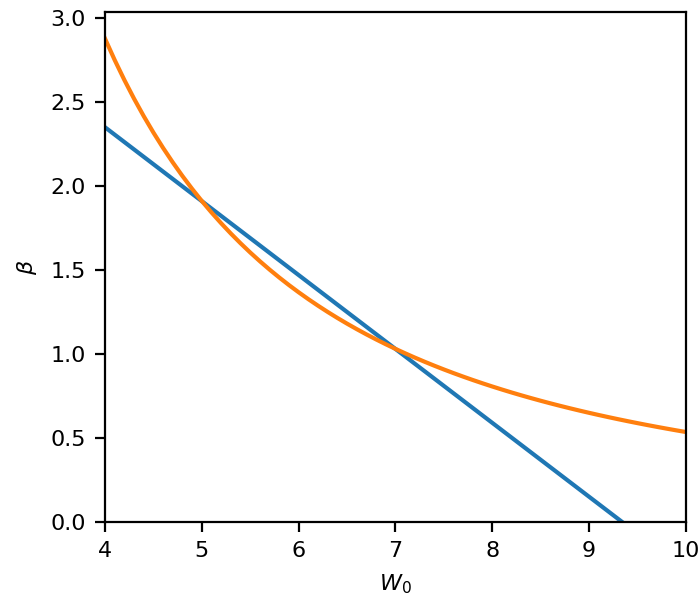}
    \caption{$\beta (W_0)$ fitted linearly (blue) and as a power-law (orange, Eq. \ref{eq_beta}).}
    \label{fig_beta}
\end{figure}

As explained in Sec. \ref{sec_ass} two important changes were made to the algorithm computing the initial masses of \acp{GC}: metallicity dependent stellar remnant masses were added and $\beta$ was computed using Eq. \ref{eq_beta} instead of a linear function.
Table \ref{tab_GCProps} shows both $M_{\rm ini}$ computed in \citet{2022MNRAS.516.3342W} and in this work next to each other.
While the computed initial masses of most \acp{GC} are lower in this work, for some they did increase.
As visible in Fig. \ref{fig_Mini}, this change is uncorrelated to the metallicities, suggesting that the change in $\beta$ is the dominant factor: Fig. \ref{fig_beta} demonstrates the difference of $\beta$, when fitted linearly compared to the new power-law (Eq. \ref{eq_beta}).

\begin{figure*}
    \sidecaption
    \includegraphics[width = 12cm]{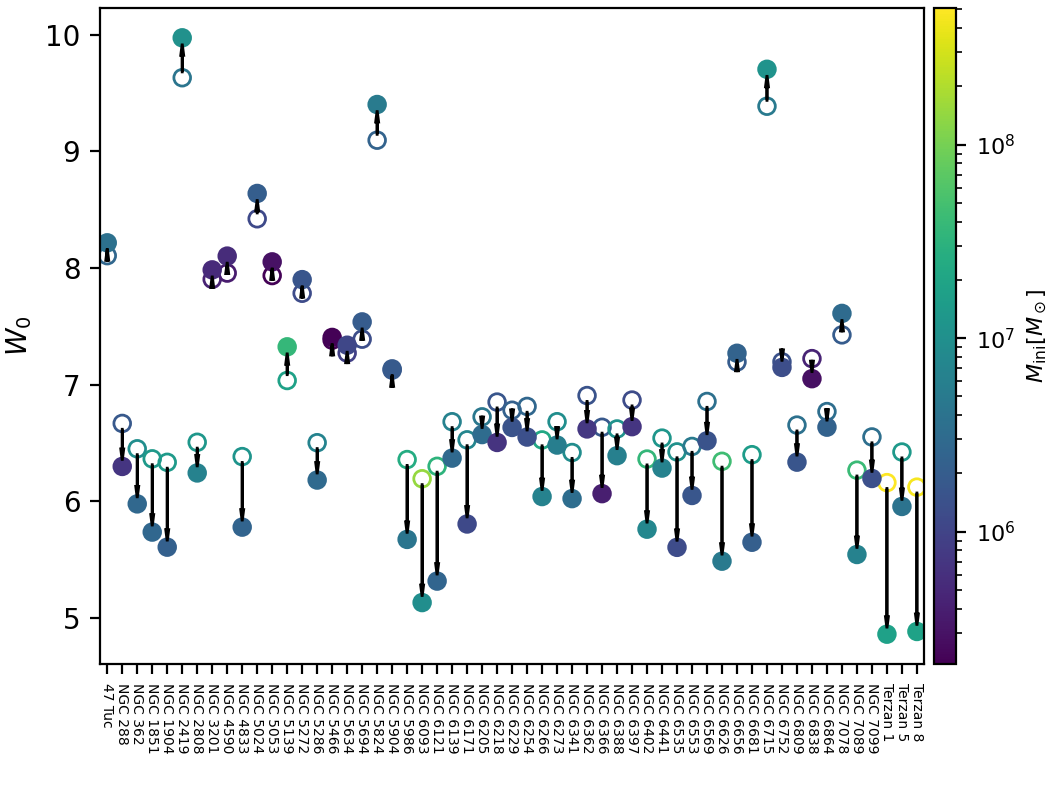}
    \caption{Old and new values for $W_0$ for the \acp{GC} in our sample.
    The values from \citet{2022MNRAS.516.3342W} are shown using an empty circle, the values for this paper using a filled circle.
    Both are colour-coded for the initial masses of the clusters and connected with an arrow pointing in the direction of the new value computed in this work.}
    \label{fig_W0}
\end{figure*}

Fig. \ref{fig_W0} depicts the King concentration parameters and initial masses of our sample used in \citet{2022MNRAS.516.3342W} and this work.
It is visible that a larger concentration parameter automatically leads to a larger initial mass, which is to be expected since the other parameters $W_0$ depends on, the pericentre radius and the \ac{SFE}, do not change between this work and \citet{2022MNRAS.516.3342W}.
Note that $r_{\rm h}$ depends on $M_{\rm ini}$ and is therefore not mentioned as a separate dependency of $W_0$.
Note also that the largest $W_0$ for the masses from \citet{2022MNRAS.516.3342W} is slightly above the the point at which the linearly fitted $\beta$ becomes negative.
The reason why this still lead to a positive $\beta$ in \citet{2022MNRAS.516.3342W} is a bug, that lead to slightly smaller $W_0$ being computed for the initial masses.
In this plot we used the correctly computed $W_0$ \citep[see][for more details]{2022MNRAS.516.3342W}.

Fig. \ref{fig_W0} also shows that for $W_0 \lesssim 7.0$ the \acp{GC} becomes less compact ($W_0$ decreases), while for $W_0 \gtrsim 7.0$ the \acp{GC} becomes more compact ($W_0$ increases) when comparing our upgraded method to the previous one.
From Fig. \ref{fig_beta} we learned that for $5.0 < W_0 < 7.0$, $\beta (W_0)$ is smaller in this paper than in \citet{2022MNRAS.516.3342W}, while it is the other way around for all other values of $W_0$.
From Eq. \ref{eq_tsf} we learned that $T_{\rm diss}$ is proportional to $\beta$ and the mass-loss rate is, therefore, inversely proportional to $\beta$.
For $5.0 < W_0 < 7.0$ ($W_0 > 7.0$) the mass-loss rate of the \ac{GC} is, therefore, decreased (increased) and therefore the computed initial mass is smaller (larger).
The fact that $W_0 = 7.0$ is not a hard boundary is due to the remnant masses now being metallicity dependent, while they were independent on the metallicity in \citet{2022MNRAS.516.3342W}.

\subsection{The upper and lower limits for SF durations}

\begin{figure}
    \includegraphics[scale=1.0]{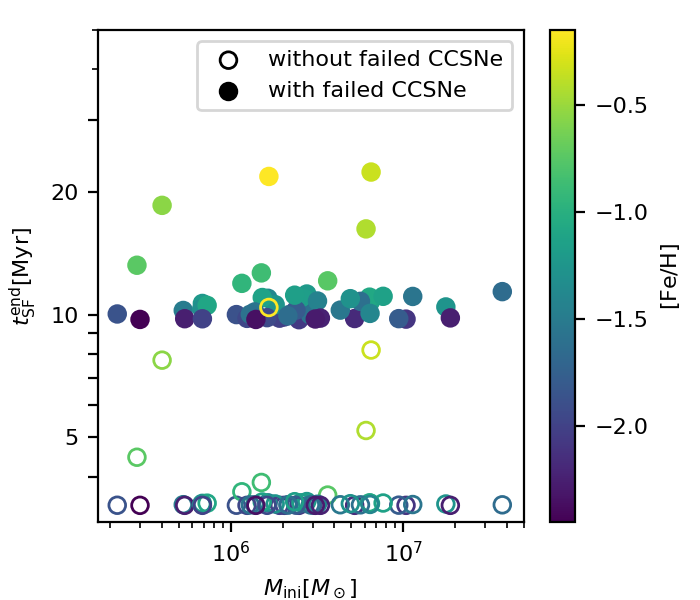}
    \caption{Time after which SF ends depending on the initial cluster mass.
    The empty circles are for the case that all stars above $8 ~\rm M_\odot$ explode in a CCSN, the filled circles show the upper limit for the SF duration, for which it is assumed that only stars with masses $< 20 ~\rm M_\odot$ explode in CCSNe.
    The plot is colour-coded for the metallicity, ${\rm [Fe/H]}$.}
    \label{fig_TSF}
\end{figure}

Table \ref{tab_GCProps} shows measured and computed quantities for the 55 \acp{GC} from \citet{2019ApJS..245....5B}.
For Terzan 5 the \ac{SF} durations could not be computed.
Even if all \acp{CCSN} that it can produce happen (all stars with a mass above $8 ~M_\odot$ explode) the iron produced is insufficient to explain the iron abundance spread observed in this \ac{GC} \citep[see also][]{2021MNRAS.506.4131W,2022MNRAS.516.3342W}.
This suggests either that this \ac{GC}, which is a bulge \ac{GC} \citep{2007AJ....133.1287V} formed while being enriched externally or that the amount of gas expelled before forming the enriched population was significantly underestimated.

According to the data from \citet{2019A&A...629A..93Y} a star with a mass of $20 ~M_\odot$ ends its life after $\approx 10.2 ~\rm Myr$.
Therefore, no explosions should happen before this time and the respective values for $t^\mathrm{SF}$ are expected to be larger than $10.2 ~\rm Myr$, which is the case.
For most \acp{GC} the upper limit remains below $12 ~\rm Myr$.
This is consistent with the findings of \citet{2013MNRAS.436.2852B} who studied 130 young massive clusters with ages between $10$ and $300 ~\rm Myr$ observationally and found no ongoing \ac{SF} within them, concluding that \ac{SF} in \acp{GC} must have ended before the age of $10 ~\rm Myr$.
This confirms the upper limit of our calculations.
However, confirming the lower limit from observational studies is more difficult.

Observational studies on very young star clusters yield vastly different results for the duration of \ac{SF}.
Estimates for NGC 3603 YC for example go from 0.4 Myr \citep{2012ApJ...750L..44K} to 10 Myr \citep{2010ApJ...720.1108B}.
The density and velocity dispersion profiles of this cluster suggest a prompt monolithic collapse of a molecular cloud clump rather than a prolonged formation from merging sub-clusters \citep{2015MNRAS.447..728B,2018ASSL..424..143B}.
For Westerlund 1 studies find 0.1 Myr \citep{2012ApJ...750L..44K} to 1 Myr \citep{2010A&A...516A..78N}.
\citet{deshmukh2024clearingtimescaleinfraredselectedstar} investigated the system of young clusters around M83 and found that for most of them the majority of the gas is cleared by pre-\ac{CCSN} feedback.
However, they also identify a massive ($10^{5.13} M_\odot$) cluster that is still surrounded by gas and dust after 6.6 Myr.
It should be pointed out here that these young clusters are more metal rich than the \acp{GC} studied in the present work, which means that the stars within them produce much stronger stellar winds \citep{2011MNRAS.415.3439D}.
This leads to an earlier gas expulsion and, therefore, a shorter duration of \ac{SF}.

The time after which \ac{SF} ends for the different \acp{GC} is visualized in Fig. \ref{fig_TSF}.
As is visible the mass of the most massive star which explodes as a \ac{CCSN} has a major impact on the time \ac{SF} ends.
Additionally, the spread between the \acp{GC} becomes larger for the cases with failed \acp{CCSN} compared to the cases where all stars above $8 ~M_\odot$ explode as a \acp{CCSN}.
This is because the decrease in stellar life expectancies with mass is lower for high-mass O stars, than for low-mass O stars \citep[see fig. 3 in][]{2019A&A...629A..93Y}.

The order in which \acp{GC} cease \ac{SF} also changes.
In Tab. \ref{tab_GCProps} we easily find pairs of \acp{GC} where one \ac{GC} has a smaller duration of \ac{SF} than the other if all stars are assumed to explode in \acp{CCSN}, but a larger duration of \ac{SF} if only stars below $20 ~M_\odot$ are allowed to explode.
This happens for example with NGC 5139 and NGC 5272 or NGC 6441 and NGC 6553.
The cause of this is the differing \acp{IMF} for the individual \acp{GC}.
A \ac{GC} with a large $M_{\rm ini}$ and low metallicity is expected to have a top-heavy \ac{IMF} \citep{2012MNRAS.422.2246M,2021A&A...655A..19Y,2022MNRAS.516.3342W}.

\begin{figure}
    \includegraphics[scale=1]{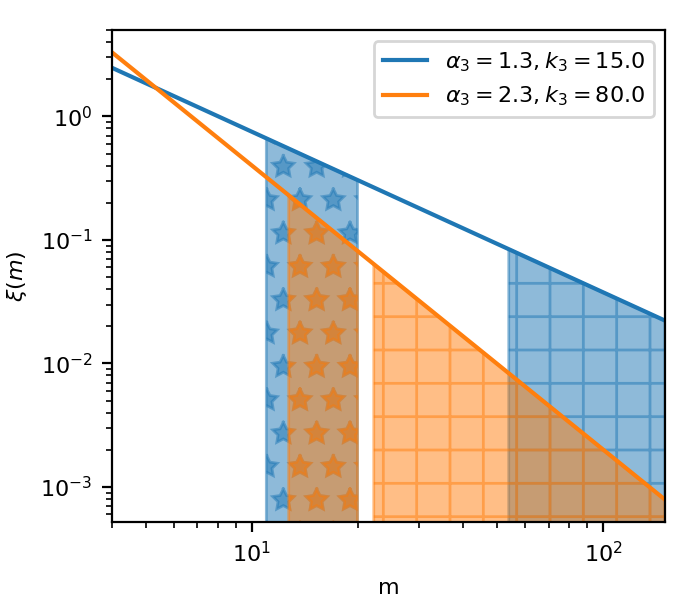}
    \caption{Example of two IMFs (one shown in blue, one in orange) with different numbers of CCSNe.
    The number of CCSNe equals the integral from the least massive star to contribute to SF to the most massive star.
    The areas below the functions, therefore, show the number of CCSNe that are expected to explode before SF ends for the case that all stars up to a mass of 150 $M_\odot$ explode in a CCSN (squares) and the case that only stars below 20 $M_\odot$ explode (stars).
    In this work the mass of the least massive star to explode (lower boundary of the coloured areas) is computed from the known number of CCSNe and the known mass of the most massive star to explode.}
    \label{fig_AreaComp}
\end{figure}

How the \ac{IMF} affects the \ac{SF} duration is demonstrated in Fig. \ref{fig_AreaComp}.
This figure shows two \acp{IMF} for two fictional stellar clusters.
For the blue cluster it is assumed that 4 \acp{CCSN} exploded before \ac{SF} ends while for the orange one it is assumed to be only one.
The coloured areas are the integral over all the stars that explode in a \ac{CCSN} before \ac{SF} ends that is they are equal to these numbers of \acp{CCSN}.
Both cases, if the most massive \ac{CCSN} progenitor has a mass of $150 ~M_\odot$ and if the most massive \ac{CCSN} progenitor has a mass of $20 ~M_\odot$, are shown.
Note that the two blue (orange) areas both have an area of 4 (1).
The low mass boundary of these areas then are equal to the masses of the last star to explode.
The more massive this star is, the shorter is the duration of \ac{SF}.
It is therefore visible how the \ac{IMF} shape and the assumptions about which stars explode in a \ac{CCSN} both affect in which \ac{GC} \ac{SF} ceases first.

The duration of \ac{SF} is especially important for the formation of multiple stellar populations in \acp{GC}, since a longer \ac{SF} duration means a prolonged pollution of the star forming gas in the \ac{GC} from stellar winds \citep{2007A&A...464.1029D,2009A&A...507L...1D,2023arXiv231010589J}.
Additionally, binaries will also experience more encounters before \ac{SF} ends if the duration of \ac{SF} is long.
This would lead to more binaries tightening and colliding, thus contributing their processed material to the surrounding gas \citep{2010MNRAS.407..277S,2020MNRAS.491..440W,2022A&A...664A.145D, 2022MNRAS.512.2936K, 2024MNRAS.527.7005K}.
It would also shorten the timescale for matter ejected from interacting binaries as discussed in \citet{2024arXiv240505687N}.
Their work investigates the evolution of binaries experiencing Roche-Lobe overflow.
This leads to a spin-up of the secondary, which then due to its quick rotation cannot accrete more material, so that the polluted material is injected into the inter-stellar gas.
However, they do not take dynamical interactions into account, therefore neglecting the effects of dynamical interactions and \acp{CCSN}.
With dynamical interactions, we would expect the Roche-Lobe overflows to happen earlier, which means that material is ejected earlier and in a less polluted state.
Therefore, an early \ac{SF} period of up to $\approx 10 ~\rm Myr$ has to be taken into account when investigating the formation of \acp{GC} and in rare cases (\acp{GC} with an unusually high amount of failed \acp{CCSN}) \ac{SF} duration can last even longer, leading to an increased enrichment with light elements.

\subsection{Precise calculations}
\label{sec_prec_calc}

To test the algorithm implemented in \ac{SFDE} we use model W18 from \citet{2016ApJ...821...38S}.
The reason we choose this model is that \citet{2016ApJ...821...38S} provides precise values for the iron ejected by the \acp{CCSN} (see their fig. 12).
The \ac{SF} duration computed using this model ($t^{\rm SF}_{\rm W18}$ in Tab. \ref{tab_GCProps}) is always larger than or equal to our lower limit ($t^{\rm SF}_{\rm all}$ in Tab. \ref{tab_GCProps}) and in most cases smaller than our upper limit ($t^{\rm SF}$ in Tab. \ref{tab_GCProps}).
For NGC 6441 and NGC 6553 the iron produced by all \acp{CCSN} exploding in the \ac{GC} is less than the amount of iron needed.
Therefore, no \ac{SF} duration could be computed for these \acp{GC}.

We also computed what happens if we assume that half of the residual gas is ejected from the cluster before iron enriched stars form ($t^{\rm SF}_{\rm W18, 0.5 M_{gas}}$ in Tab. \ref{tab_GCProps}).
This effectively halves the amount of iron required to explain the iron spread, reducing the number of \acp{CCSN} required to produce the iron.
For most \acp{GC} this significantly decreases the duration of \ac{SF}.
However, for \acp{GC} with already very small durations of \ac{SF} it stays constant.
This is caused by the low variations of stellar lifetimes in high-mass stars (no variation for the highest mass stars in this work).

\begin{figure}
    \subfigure[]{\includegraphics[scale=1]{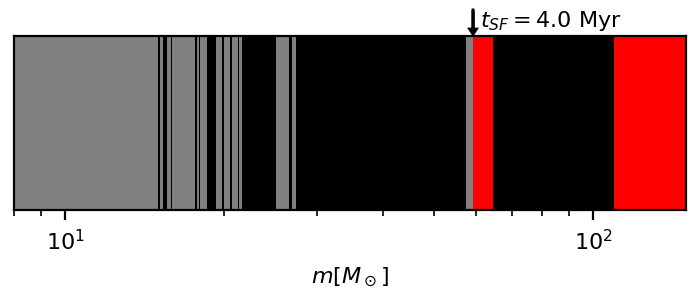}}
    \subfigure[]{\includegraphics[scale=1]{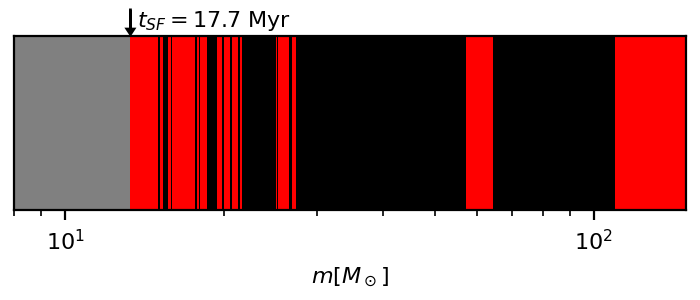}}
    \caption{Outcomes of stars depending on their stellar initial mass for NGC 288 and NGC 6366 using the CCSN model W18 of \citet{2016ApJ...821...38S}.
    Black shows mass ranges of failed CCSNe, red shows stars that explode as CCSNe and of which the iron is used in the formation of further stars and grey shows stars that explode as CCSNe after SF stopped.
    The time after which SF ends is marked by an arrow and written above the last star to explode.}
    \label{fig_barcodes}
\end{figure}

Fig. \ref{fig_barcodes} is designed similar to the `barcodes' shown in fig. 13 of \citet{2016ApJ...821...38S}.
The stars that contribute iron to the formation of new stars are marked in red.
As is visible, a few of the stars with masses between $8$ and $20 ~\rm M_\odot$ still end in failed \acp{CCSN} and some of the stars with masses above $20 ~\rm M_\odot$ explode as \acp{CCSN}.
Additionally, since the estimates of ejecta masses from \citet{2016ApJ...821...38S} are used, some stars produce a different amount from the $0.074 ~M_\odot$ assumed previously for computing the upper and lower limit.

\begin{figure}
    \centering
    \subfigure[]{\includegraphics[scale=1]{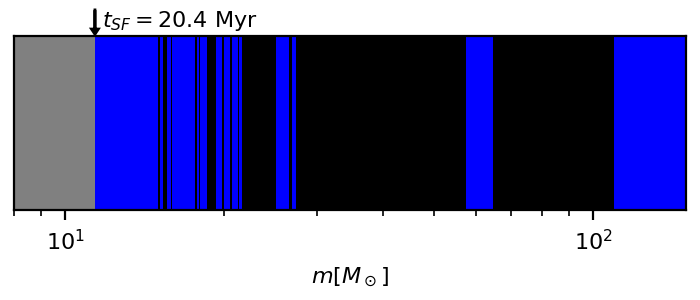}}
    \subfigure[]{\includegraphics[scale=1]{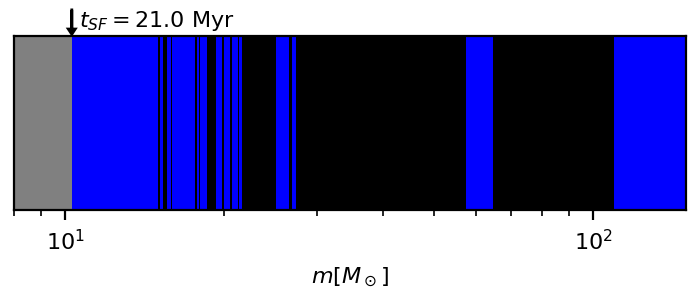}}
    \caption{Outcomes of stars depending on their initial stellar mass in NGC 6441 (a) and NGC 6553 (b) using the failed CCSNs of model W18 of \citet{2016ApJ...821...38S}, but a constant amount of $0.074 ~M_\odot$ of iron ejected per CCSN.
    Black shows mass ranges of failed CCSNe, blue shows stars that explode as CCSNe and of which the iron is used in the formation of further stars and grey shows stars that explode as CCSNe after SF stopped.
    The time after which SF ends is marked by an arrow and written above the last star to explode.}
    \label{fig_barcodes2}
\end{figure}

\begin{figure}
    \centering
    \subfigure[]{\includegraphics[scale=1]{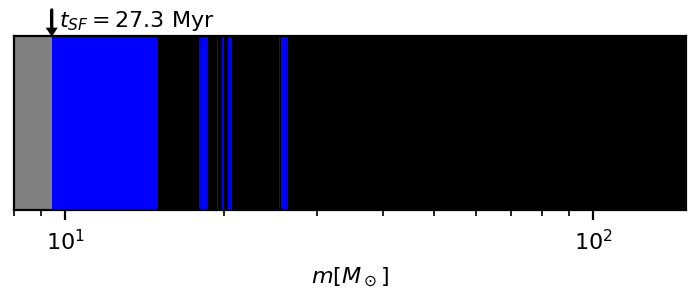}}
    \subfigure[]{\includegraphics[scale=1]{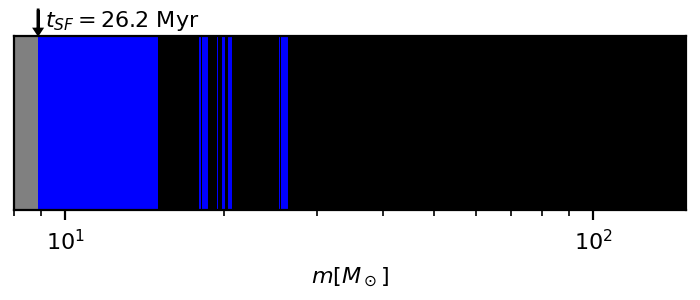}}
    \caption{Same as Fig. \ref{fig_barcodes2}, but model W20 from \citet{2016ApJ...821...38S} is used instead of W18.}
    \label{fig_barcodes3}
\end{figure}

To understand the influence these two effects have on the \ac{SF} duration, the calculations for the two \acp{GC} that couldn't produce enough iron in total (NGC 6441 and NGC 6553) were done again with the same mass ranges for failed \acp{CCSN}, but assuming that each \ac{CCSN} produces $0.074 ~M_\odot$ of iron.
The results are shown in Fig. \ref{fig_barcodes2}, with blue used for the stars contributing to the iron spread to underline the difference in assumed masses of iron produced.
Both NGC 6441 and NGC 6553 now have \ac{SF} durations within the computed lower and upper limit, showing that the difference in iron output between the models has a significant effect on the \ac{SF} duration.
As Fig. \ref{fig_barcodes3} shows, even redoing the same calculations with Model W20 from \citet{2016ApJ...821...38S}, the most extreme model (when it comes to the ranges of failed \acp{CCSN}) available to us, still produces enough iron to explain the observed iron spread.
However the \ac{SF} duration now exceeds our previously computed upper limit.
Therefore, the determination of the \ac{SF} duration depends strongly on the precise iron ejecta of \ac{CCSN} assumed for the calculation.

\section{Discussion}
\label{sec_disc}

\subsection{Discussion of the assumptions}

In this work we improved the model from \citet{2022MNRAS.516.3342W} by adding the effects of failed \acp{CCSN} and varying the amount of iron ejected per star.
However, many aspects of this problem still require further investigation.

In our model we assumed that all ejecta were preserved in the gas cloud and that the gas would be well mixed.
As described in \citet{2015ApJ...814L..14C} the ejecta from \acp{CCSN} cuts tunnels into the remaining gas cloud, through which gas can escape the \ac{GC} easier.
A loss of some of the \ac{CCSN} ejecta would lead to a higher number of \acp{CCSN} required to produce the observed iron spread.
On the other hand the \acp{CCSN} would produce pockets of higher metallicity, therefore, reducing the number of \ac{CCSN} required.
It is unclear, which of the two effects is dominant.

Furthermore, any kind of accretion and binary evolution is neglected.
Both have been proposed as a possible cause for the multiple population phenomenon \citep{2009A&A...507L...1D,2010MNRAS.407..277S,2018MNRAS.478.2461G,2020MNRAS.491..440W,2022A&A...664A.145D,2022MNRAS.512.2936K,2024MNRAS.527.7005K,2024arXiv240505687N}.
Observational studies have shown that accretion onto a low-metallicity star can go on for up to $10 ~\rm Myr$ \citep{2010A&A...510A..72F}, which means that previously non-enriched stars may experience a small enrichment from polluted material entering the accretion disk.
The magnitude of this effect is currently unknown.
Additionally, stars do also show chemical enrichment, when ingesting a planet \citep{2024arXiv240313209L}.
These stars might become visible as exotic stars in a stellar cluster.
In the current study both of these effects are neglected due to their unknown extent.
Binary evolution would also greatly affect the outcomes of stellar evolution due to changes of the stellar mass caused by mass transfer and mergers.
It is currently unclear what kind of ejecta are produced this way.

\subsection{The current understanding of failed \acp{CCSN}}

As mentioned in Sec. \ref{sec_intr} it is not yet fully understood under which conditions massive stars explode as \acp{CCSN} and under which conditions they implode into a \ac{BH}.
Additionally, for those stars that do explode, the amount of iron ejected is not uniform.
\cite{1998A&A...334..505P} describe the composition of \acp{CCSN} depending on the initial masses and metallicities of stars assuming all stars above $8 ~M_\odot$ explode.
They find variations in iron outputs of about 3 orders of magnitude with a minimum iron output for stars with an initial mass of around $30 ~M_\odot$.

Newer studies, however, describe the possibility of failed \acp{CCSN}.
Additionally, if an explosion occurs, the nature of the remnant and, therefore, the ejecta is not only dependent on mass and metallicity.
\citet{2011ApJ...730...70O} point out that rotation significantly affects a \ac{CCSN} due to centrifugal forces.
Similarly \citet{2018ApJS..237...13L} found larger ejecta masses in fast rotating stars when compared to their slower or non-rotating counterparts.
\citet{2016ApJ...821...38S} point out the dependency on the structure of the pre-\ac{CCSN} core, while \citet{2015ApJ...801...90P} find a strong dependency on the assumed required neutrino luminosity, showing how large the uncertainties are in current models of \acp{CCSN}.
Finally, \citet{2019ApJ...870....1E} confirm the dependency on rotation and core structure and point out that the magnetic field of the star also affects the outcome of stellar evolution.

It is currently unknown how many stars and stars of which masses and metallicities actually end up exploding as a \ac{CCSN} and what the exact amount and composition of the ejecta are.
As was shown, this has a large effect on our results.
To make more precise deductions about the duration of \ac{SF} during \ac{GC} formation, reliable \acp{CCSN} models for stars depending on metallicity, initial mass and rotation are needed.
Furthermore, the initial distribution of stellar rotation in a just-born \ac{GC} needs to be investigated.

\section{Conclusions}
\label{sec_concl}

This work revisits the results of \citet{2022MNRAS.516.3342W} and studies the upper limit for the \ac{SF} durations of \acp{GC} assuming that all stars more massive than $20 ~M_\odot$ result in failed \acp{CCSN}.
The lower limit computed in \citet{2022MNRAS.516.3342W} under the assumption that all stars above $8 ~M_\odot$ explode as \acp{CCSN} ($\approx 3.5 ~\rm Myr$) and the upper limit computed here ($\approx 10.5 ~\rm Myr$) differ by about a factor of 3 which has significant implications on the formation of \acp{GC}.
It is also shown that failed \acp{CCSN} significantly affect the order in which \acp{GC} born simultaneously seize star formation.
Because of this no conclusions about the duration of \ac{SF} in different \acp{GC} relative to one another can be made without a more precise understanding of which \acp{CCSN} fail.
However, as mentioned before, the longer durations of \ac{SF} allow for longer periods of pollution from binaries and stellar winds to form multiple populations \citep{2009A&A...507L...1D,2010MNRAS.407..277S,2018MNRAS.478.2461G,2020MNRAS.491..440W,2022A&A...664A.145D,2022MNRAS.512.2936K,2024MNRAS.527.7005K,2024arXiv240505687N}.

To further improve our understanding of the duration of \ac{SF}, the nature of \ac{CCSN} and the dependence of the amount of the ejected iron on stellar initial mass, metallicity and rotation needs to be investigated further.
It is also important to develop a better understanding of the stellar rotation distribution in the young \acp{GC} as this largely influences the amount and chemical composition of ejecta produced \citep{2011ApJ...730...70O,2018ApJS..237...13L}.

\begin{acknowledgements}
We thank an anonymous referee for many helpful suggestions.
We also thank the DAAD-Eastern-European exchange programme at Bonn and Charles University for support.
\end{acknowledgements}



\bibliographystyle{aa}
\bibliography{IronSpreadGC2} 

\begin{appendix}
\onecolumn
\section{Additional table}

\begin{table*}[h]
    \caption{The deduced masses, metallicities, numbers of CCSNe and times when SF ends for 55 Galactic GCs.
    }
    \begin{center}
    \begin{small}
        \begin{tabular}{lrrrrrrrr}\hline
			Name & \multicolumn{1}{c}{$M_{\rm ini}~[10^5 M_\odot]$} & \multicolumn{1}{c}{$M_{\rm ini}^{\rm old}~[10^5 M_\odot]$} & {${\rm [Fe/H]}$} & \multicolumn{1}{c}{$N_{\rm SN}$} & \multicolumn{1}{c}{$t_{\rm all}^{\rm SF}~[{\rm Myr}]$} & \multicolumn{1}{c}{$t^{\rm SF}$} & \multicolumn{1}{c}{$t^{\rm SF}_{\rm W18}~[{\rm Myr}]$} & \multicolumn{1}{c}{$t^{\rm SF}_{\rm W18, 0.5 M_{gas}}~[{\rm Myr}]$} \\
			\hline
			47 Tuc&      36.55&    27.50&   -0.747&{$4.31\times 10^{3}$}&      3.6&     12.1&    7.8&    4.1\\
			NGC 288&      6.86&    17.79&   -1.226&{$3.01\times 10^{2}$}&      3.4&     10.7&    4.0&    3.4\\
			NGC 362&     27.66&    93.75&   -1.213&{$2.51\times 10^{3}$}&      3.5&     11.2&    7.3&    3.9\\
			NGC 1851&    26.92&   136.44&   -1.157&{$1.72\times 10^{3}$}&      3.5&     10.9&    4.0&    3.4\\
			NGC 1904&    21.68&   142.09&   -1.550&{$3.29\times 10^{2}$}&      3.4&     10.0&    3.4&    3.4\\
			NGC 2419&   104.06&    42.97&   -2.095&{$5.34\times 10^{2}$}&      3.4&      9.8&    3.4&    3.4\\
			NGC 2808&    64.73&   127.73&   -1.120&{$3.43\times 10^{3}$}&      3.4&     10.8&    3.9&    3.4\\
			NGC 3201&     5.32&     4.32&   -1.496&{$1.49\times 10^{2}$}&      3.4&     10.2&    3.8&    3.4\\
			NGC 4590&     5.42&     3.72&   -2.255&{$3.19\times 10^{1}$}&      3.4&      9.8&    3.4&    3.4\\
			NGC 4833&    24.91&   118.88&   -2.070&{$5.50\times 10^{1}$}&      3.4&      9.7&    3.4&    3.4\\
			NGC 5024&    20.49&    11.69&   -1.995&{$2.95\times 10^{2}$}&      3.4&      9.9&    3.4&    3.4\\
			NGC 5053&     2.98&     2.20&   -2.450&{$8.67\times 10^{0}$}&      3.4&      9.7&    3.4&    3.4\\
			NGC 5139&   376.26&   178.36&   -1.647&{$4.89\times 10^{4}$}&      3.4&     11.4&    7.0&    3.4\\
			NGC 5272&    16.46&    12.17&   -1.391&{$1.31\times 10^{3}$}&      3.5&     11.0&    7.2&    3.9\\
			NGC 5286&    29.91&    68.41&   -1.727&{$1.16\times 10^{3}$}&      3.4&     10.2&    3.4&    3.4\\
			NGC 5466&     2.20&     2.07&   -1.865&{$4.52\times 10^{1}$}&      3.4&     10.0&    3.8&    3.4\\
			NGC 5634&    10.81&     9.09&   -1.869&{$2.38\times 10^{2}$}&      3.4&     10.0&    3.4&    3.4\\
			NGC 5694&    19.21&    13.10&   -2.017&{$1.70\times 10^{2}$}&      3.4&      9.8&    3.4&    3.4\\
			NGC 5824&    52.40&    23.77&   -2.174&{$4.07\times 10^{2}$}&      3.4&      9.8&    3.4&    3.4\\
			NGC 5904&    18.13&    17.63&   -1.259&{$8.18\times 10^{2}$}&      3.4&     10.6&    3.9&    3.4\\
			NGC 5986&    43.26&   253.81&   -1.527&{$1.57\times 10^{3}$}&      3.4&     10.3&    3.4&    3.4\\
			NGC 6093&    94.42&  1467.87&   -1.789&{$4.29\times 10^{2}$}&      3.4&      9.8&    3.4&    3.4\\
			NGC 6121&    25.57&   325.37&   -1.166&{$1.74\times 10^{3}$}&      3.5&     11.0&    4.1&    3.4\\
			NGC 6139&    28.99&    64.92&   -1.593&{$4.88\times 10^{2}$}&      3.4&     10.0&    3.4&    3.4\\
			NGC 6171&    11.62&    75.02&   -0.949&{$1.23\times 10^{3}$}&      3.7&     11.9&    7.8&    4.1\\
			NGC 6205&    31.91&    47.26&   -1.443&{$2.34\times 10^{3}$}&      3.4&     10.8&    4.0&    3.4\\
			NGC 6218&     6.80&    16.64&   -1.315&{$1.91\times 10^{2}$}&      3.4&     10.3&    3.9&    3.4\\
			NGC 6229&    15.26&    22.35&   -1.129&{$9.98\times 10^{2}$}&      3.5&     11.0&    7.2&    3.9\\
			NGC 6254&    13.82&    27.25&   -1.559&{$3.74\times 10^{2}$}&      3.4&     10.2&    3.4&    3.4\\
			NGC 6266&    63.96&   224.31&   -1.075&{$4.41\times 10^{3}$}&      3.5&     11.0&    4.0&    3.5\\
			NGC 6273&    56.62&    95.00&   -1.612&{$4.55\times 10^{3}$}&      3.4&     10.8&    3.9&    3.4\\
			NGC 6341&    33.17&    92.11&   -2.239&{$3.19\times 10^{2}$}&      3.4&      9.8&    3.4&    3.4\\
			NGC 6362&     7.31&    15.32&   -1.092&{$2.01\times 10^{2}$}&      3.4&     10.5&    3.9&    3.4\\
			NGC 6366&     4.01&    17.45&   -0.555&{$1.59\times 10^{3}$}&      7.7&     18.6&   17.7&   12.6\\
			NGC 6388&    61.01&   110.26&   -0.428&{$2.46\times 10^{4}$}&      5.2&     16.2&   14.7&   10.3\\
			NGC 6397&     6.86&    12.37&   -1.994&{$3.89\times 10^{1}$}&      3.4&      9.8&    3.4&    3.4\\
			NGC 6402&    76.64&   362.45&   -1.130&{$6.02\times 10^{3}$}&      3.5&     11.1&    4.0&    3.4\\
			NGC 6441&    65.26&   126.76&   -0.334&{$4.79\times 10^{4}$}&      8.2&     22.4&    {-}&   13.6\\
			NGC 6535&    12.45&   103.20&   -1.963&{$9.48\times 10^{1}$}&      3.4&      9.8&    3.4&    3.4\\
			NGC 6553&    16.67&    49.39&   -0.151&{$1.11\times 10^{4}$}&     10.4&     21.9&    {-}&   14.0\\
			NGC 6569&    15.08&    36.03&   -0.867&{$2.25\times 10^{3}$}&      3.9&     12.7&   10.0&    7.6\\
			NGC 6626&    49.46&   452.45&   -1.287&{$3.84\times 10^{3}$}&      3.4&     11.0&    4.0&    3.4\\
			NGC 6656&    23.94&    19.75&   -1.803&{$1.01\times 10^{3}$}&      3.4&     10.3&    3.8&    3.4\\
			NGC 6681&    21.49&   149.57&   -1.633&{$2.80\times 10^{2}$}&      3.4&      9.9&    3.4&    3.4\\
			NGC 6715&   113.80&    50.19&   -1.559&{$1.18\times 10^{4}$}&      3.4&     11.1&    7.0&    3.4\\
			NGC 6752&    13.00&    14.72&   -1.583&{$2.31\times 10^{2}$}&      3.4&     10.0&    3.4&    3.4\\
			NGC 6809&    16.27&    36.99&   -1.934&{$1.70\times 10^{2}$}&      3.4&      9.8&    3.4&    3.4\\
			NGC 6838&     2.86&     4.46&   -0.736&{$4.10\times 10^{2}$}&      4.5&     13.2&   11.5&    7.9\\
			NGC 6864&    23.52&    33.22&   -1.164&{$1.90\times 10^{3}$}&      3.5&     11.2&    7.3&    3.9\\
			NGC 7078&    30.98&    19.26&   -2.287&{$1.70\times 10^{2}$}&      3.4&      9.8&    3.4&    3.4\\
			NGC 7089&    64.36&   417.15&   -1.399&{$1.08\times 10^{3}$}&      3.4&     10.1&    3.4&    3.4\\
			NGC 7099&    14.02&    35.18&   -2.356&{$4.57\times 10^{1}$}&      3.4&      9.7&    3.4&    3.4\\
			Terzan 1&   177.07&  5089.44&   -1.263&{$7.14\times 10^{3}$}&      3.4&     10.5&    3.4&    3.4\\
			Terzan 5&    40.85&   136.44&   -0.092&{$2.10\times 10^{5}$}&     {-} &     {-} &    {-}&    {-}\\
			Terzan 8&   188.04&  4604.79&   -2.255&{$2.06\times 10^{3}$}&      3.4&      9.8&    3.4&    3.4\\
			\hline
	    \end{tabular}
    \end{small}
        \label{tab_GCProps}
    \end{center}
    \tablefoot{The columns from left to right are: the name of the GC, the initial mass, $M_{\rm ini}$, the initial mass computed by \citet{2022MNRAS.516.3342W}, $M_{\rm ini}^{\rm old}$, the metallicity, ${\rm [Fe/H]}$, the number of CCSNe, $N_{\rm SN}$, the time SF ends assuming: 1. all stars above $8 M_\odot$ explode as CCSNe, $t^{\rm SF}_{\rm all}$, 2. all CCSNe of stars above $20 M_\odot$ fail, $t^{\rm SF}$, 3. model W18 from \citet{2016ApJ...821...38S} and 4. model W18 from \citet{2016ApJ...821...38S} with only half of the total gas mass left.}
    \vspace{-1.5cm}
\end{table*}

\end{appendix}

\end{document}